 \documentclass[manuscript]{emulateapj}

 \usepackage{apjfonts}

\slugcomment{accepted on 12th September 2011; to appear in the Astrophysical Journal}

\shorttitle{Super Chandrasekhar Mass Model}
\shortauthors{Hachisu et al.}

\begin{document}

\title{A Single Degenerate Progenitor Model for Type I\lowercase{a}
Supernovae Highly Exceeding the Chandrasekhar Mass Limit}


\author{Izumi Hachisu}
\affil{Department of Earth Science and Astronomy,
College of Arts and Sciences, University of Tokyo,
Komaba 3-8-1, Meguro-ku, Tokyo 153-8902, Japan}
\email{hachisu@ea.c.u-tokyo.ac.jp}

\author{Mariko Kato}
\affil{Department of Astronomy, Keio University,
Hiyoshi 4-1-1, Kouhoku-ku, Yokohama 223-8521, Japan}
\email{mariko@educ.cc.keio.ac.jp}

\author{Hideyuki Saio}
\affil{Astronomical Institute, Graduate School of Science, Tohoku University,
Sendai 980-8578, Japan}
\email{saio@astr.tohoku.ac.jp}

\and

\author{Ken'ichi Nomoto}
\affil{
Institute for the Physics and Mathematics of the Universe,
University of Tokyo, Kashiwa, Chiba 277-8583, Japan}
\email{nomoto@astron.s.u-tokyo.ac.jp}



\begin{abstract}
Recent observations of Type~Ia supernovae (SNe~Ia) suggest that
some of the progenitor white dwarfs (WDs) had masses
up to 2.4--$2.8~M_\sun$, highly exceeding the Chandrasekhar mass limit.  
We present a new single degenerate (SD) model for SN~Ia
progenitors, in which the WD mass possibly reaches 2.3--$2.7~M_\sun$.
Three binary evolution processes are incorporated;
optically thick winds from mass-accreting WDs, mass-stripping
from the binary companion star by the WD winds, and WDs being
supported by differential rotation.  The WD mass can increase
by accretion up to 2.3~(2.7)$~M_\sun$ from the initial value
of 1.1~(1.2)$~M_\sun$, being consistent with high luminosity SNe~Ia
such as SN~2003fg, SN~2006gz, SN~2007if, and SN~2009dc.
There are three characteristic mass ranges of exploding WDs.
In an extreme massive case, differentially rotating WDs explode
as an SN~Ia soon after the WD mass exceeds $2.4~M_\sun$
because of a secular instability at $T/|W|\sim 0.14$.
For a mid mass range of $M_{\rm WD}=1.5$--$2.4~M_\sun$,
it takes some time (spinning-down time)
until carbon is ignited to induce an SN Ia explosion
after the WD mass has reached maximum, because it needs
a loss or redistribution of angular momentum.
For a lower mass case of rigidly rotating WDs, 
$M_{\rm WD}=1.38$--$1.5~M_\sun$, the spinning-down time
depends on the timescale of angular momentum loss from the WD.
The difference in the spinning-down time may produce
the ``prompt'' and ``tardy'' components.
We also suggest the very bright super-Chandrasekhar mass SNe~Ia
are born in a low metallicity environment.
\end{abstract}




\section{Introduction}
Type Ia supernovae (SNe~Ia) play important roles
in astrophysics, e.g., standard candles in cosmological distance
and main production cites of iron-group elements.  However,
the nature of SN~Ia progenitors is not fully understood yet 
\citep[e.g.,][]{nie04, nom00}.
It has been widely accepted that SNe~Ia are explosions of
carbon-oxygen (C+O) white dwarfs (WDs).
The observed features of SNe~Ia are better explained
by the Chandrasekhar mass model rather than the sub-Chandrasekhar
mass model. It is not clear, however, 
how the WD mass gets close enough to the Chandrasekhar
mass\footnote{Hereafter we use the Chandrasekhar mass $M_{\rm Ch}=1.46
(Y_{\rm e}/0.5)^2M_\sun$ \citep[e.g.,][]{cha39} for a
{\it non-rotating} WD with electron mole number $Y_{\rm e}$.}
for carbon ignition; i.e., whether the WD accretes H/He-rich matter from
its binary companion [single degenerate (SD) scenario], or two C+O WDs
merge [double degenerate (DD) scenario].

Recent observations of several very bright SNe~Ia suggest
that their progenitor WDs might have super-Chandrasekhar masses
of up to 2.4--$2.8~M_\sun$ \citep[e.g.,][]{hic07, how06, sca10, sil11,
tau11, yam09}.  Super-Chandrasekhar mass of the progenitors of
SNe~Ia was first reported by \citet{how06} on SN~2003fg 
($\sim 2.1~M_\sun$ C+O WD with $1.29 \pm 0.07~M_\sun~ ^{56}$Ni).
More candidates of super-Chandrasekhar mass progenitors were
added to the list, i.e., SN~2006gz, SN~2007if, and SN~2009dc.
\citet{hil07} suggested a possibility that SN~2003fg is an
aspherical explosion of a Chandrasekhar mass WD.  However, 
based on their recent spectropolarimetric observation of
SN~2009dc which shows little continuum polarization of $<0.3$
\citet{tan10} concluded that a very large asphericity 
was not likely to occur at the explosion of SN~2009dc.  
Moreover, the very bright SN~2007if 
require a $2.4\pm0.2~M_\sun$ progenitor C+O WD with
$1.6\pm0.1~M_\sun$ $^{56}$Ni \citep{sca10} and SN~2009dc also
demands $\gtrsim 2.0~M_\sun$ progenitor C+O WD with
1.4--1.7$~M_\sun$ $^{56}$Ni \citep{sil11} 
or $\sim 2.8~M_\sun$ C+O WD with $\sim 1.8~M_\sun$ $^{56}$Ni \citep{tau11}.
Such a large WD mass, highly exceeding $M_{\rm Ch}$, could
challenge both the DD and SD scenarios.
Some authors suggest a DD merger scenario for
these super-Chandrasekhar mass SNe~Ia \citep[e.g.,][]{sca10, sil11},
mainly because the existing SD
scenario provides only $M_{\rm WD}\lesssim1.8~M_\sun$
\citep[e.g.,][]{che09,liu10}.

As for theoretical explosion models of super-Chandrasekhar mass WDs,
\citet{ste92} performed simulations of pure detonation models and
found that almost all the matter is burned into iron group elements.
Their results could not explain the observed intermediate mass elements
seen in the spectra of SNe~Ia.  Recently, \citet{pfa10} revised
the simulations of pure detonation models based on Yoon \& Langer's
(2005) super-Chandrasekhar mass models.  They found that not all the
WD matter is burned into iron group elements and some intermediate mass
elements are synthesized in the outer low density region of the WD.
The synthesized mass of nickel is large enough ($\sim 1.5~M_\sun$)
to explain the super luminous SNe~Ia such as SN~2009dc, which  supports
the super-Chandrasekhar mass models.

  In the present paper, we show that such high WD masses are possibly
explained by the SD scenario. There are three key binary evolution
processes, i.e., optically thick winds from mass-accreting WDs
\citep{hkn96}, mass-stripping from the binary companion star
by the WD winds \citep{hkn08a}, and differential rotation which makes
possible a WD more massive than the Chandrasekhar limit \citep{yoo04}.
The main difference of our SD model from the existing SD scenarios is
that we take into account the effect of mass-stripping by the winds
from mass-accreting WDs.  We describe our basic assumptions and methods
of binary evolutions in Section 2.  Our numerical results are presented
in Section 3.  A possible binary evolution to our super-Chandrasekhar
mass SNe~Ia is suggested in Section 4.  Discussion and concluding remarks
follow in Sections 5 and 6, respectively.

\section{Modelling of Mass-Stripping Effect and Binary Evolution}
First we briefly describe the method of
our calculations for binary evolutions which is essentially
the same as in \citet{hkn08a} except that the evolution of
the companion star is computed by a Henyey-type code.  
In a binary, the more massive (primary) component evolves to
a red giant star (with a helium core) or an asymptotic giant
branch (AGB) star (with a C+O core) and fills its Roche lobe.
Dynamically unstable mass transfer begins from the primary
to the secondary and forms a common envelope.
After this first common envelope evolution,
the separation shrinks and the primary
becomes a helium star or a C+O WD.  The helium star
evolves to a C+O WD after a large part of helium is exhausted
by core- and shell-helium-burning.  Then the binary consists of
a close pair of a C+O WD and a main-sequence (MS) star;
from such a pair we start our calculations. 

When the secondary expands to fill its Roche lobe, mass transfer
begins from the secondary to the WD.  If the secondary mass is
more massive than the WD (precisely, when the mass ratio 
$q=M_{2,0}/M_{1,0}=M_{\rm MS, 0}/M_{\rm WD,0}\gtrsim0.8$),
the mass transfer takes place on a thermal timescale.
When the mass transfer rate exceeds the critical rate, 
\begin{equation}
\dot M_{\rm cr}=6.68\times10^{-7}\left({{M_{\rm WD}}\over
{M_\sun}}-0.445\right)M_\sun{\rm ~yr}^{-1},
\label{steady_accretion_wind}
\end{equation}
optically thick winds blow from the WD
\citep[see][for more details]{hkn96, hkn99, hknu99, nom07}.

The winds from the WD collide with the secondary
and strips off its surface layer \citep{hac03kb, hac03kc}.
This mass-stripping effect attenuates the mass transfer from
the secondary to the WD, thus preventing the formation of
a common envelope even for a rather massive secondary.
Thus the mass-stripping effect widens the donor mass range of
SN~Ia progenitors \citep[e.g.,][]{hkn08a, hkn08b}.

We have incorporated the mass-stripping effect
in the same way as in \citet{hkn08a}; i.e.,
the mass stripping rate ${\dot M}_{\rm strip}$ is related to
the wind mass-loss rate $\dot M_{\rm wind}$ as
\begin{equation}
{\dot M}_{\rm strip}=c_1{\dot M}_{\rm wind},
\label{mass_stripping_rate}
\end{equation}
with the factor $c_1$ defined as
\begin{equation}
c_1\equiv{{\eta_{\rm eff}\cdot g(q)}\over
{\phi_{\rm L3}-\phi_{\rm MS}}}
\left({{v^2 a}\over{2 G M}}\right),
\end{equation}
where $M=M_{\rm WD}+M_{\rm MS}$, $M_{\rm WD}$ is the WD mass,
$M_{\rm MS}$ is the main-sequence companion mass,
$a$ is the separation of the binary;
$\phi_{\rm L3}$ and $\phi_{\rm MS}$ denote
the Roche potential (normalized by $GM/a$)
at the MS surface and the L3 point near the MS companion, respectively,
i.e., $\phi_{\rm L3}-\phi_{\rm MS}\sim0.3$;
$\eta_{\rm eff}$ is the efficiency of conversion from kinetic energy
to thermal energy by the shock, $g(q)\sim0.025$ is the geometrical factor
of the MS surface hit by the wind including the inclination 
(oblique shock) effect of
the wind velocity against the companion's surface
\citep[see][for more details on $g(q)$]{hkn99},
and $q\equiv M_2/M_1=M_{\rm MS}/M_{\rm WD}$
is the mass ratio.  Here we modify
equation (21) of \citet{hkn99} to include the effect of Roche lobe
overflow from the L3 point.
We assume $\eta_{\rm eff}=1$ in the present calculation.
If the wind velocity is as fast as 4,000 km~s$^{-1}$,
we have $c_1\sim10$ as estimated by \citet{hac03kb}.
\citet{hac03kb, hac03kc} found best fit models at $c_1=1.5-10$ for
RX~J0513.9$-6953$ and at $c_1=7-8$ for V~Sge, respectively.
Here we adopt $c_1= 1$, 3, and 10 to check
the dependence of mass-stripping effect on the parameter $c_1$.

\begin{figure*}
\epsscale{1.0}
\plotone{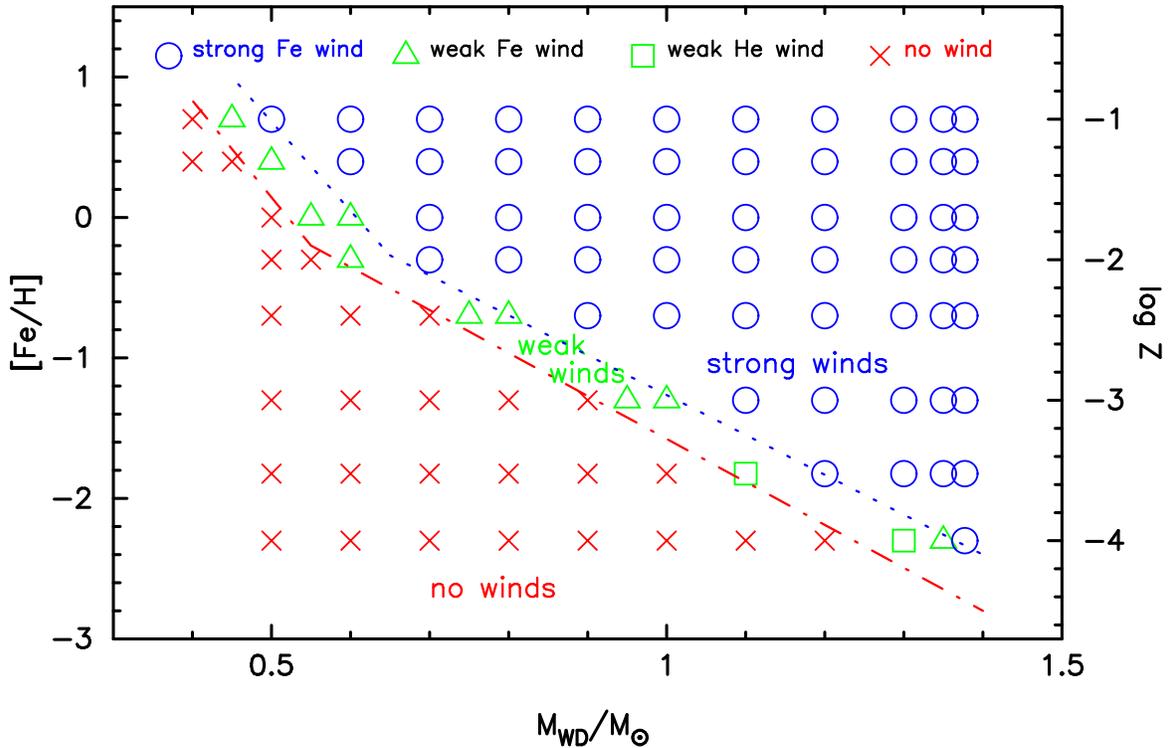}
\caption{
Wind region in the metallicity-WD mass diagram.
Data are essentially the same as those in Figure 1 of \citet{kob98}
but a few points are revised.
Each symbol shows as follows: strong winds ({\it blue circles}),
weak winds ({\it green triangles}), both of which are driven
by the OPAL peak at $\log T~({\rm K})=5.2$,
weak winds  ({\it green squares}) which are driven by the opacity
peak due to helium ionization,
and no winds ({\it red crosses}).  ``Strong winds'' mean that
the wind velocity exceeds the escape velocity while
``weak winds'' do not exceed.  The blue dotted line shows a border 
of strong winds.  The red dash-dotted line indicates a border 
of no winds.  In the low metallicity region, strong winds blow only
for very massive WDs (e.g., $M_{\rm WD} \gtrsim 1.0~M_\sun$
for [Fe/H]$\lesssim -1$ and $M_{\rm WD} \gtrsim 1.2~M_\sun$
for [Fe/H]$\lesssim -2$).
\label{metal_new}}
\end{figure*}

The winds depend on the WD mass and metallicity (chemical composition).
We have revised metallicity effects of WD winds
\citep[Figure 1 of][]{kob98}.  
The results are shown in Figure \ref{metal_new}.
We did not find wind solutions but instead static envelope solutions
for the ``no winds'' region.  On the other hand, the ``strong winds''
region means that the wind velocity exceeds the escape velocity
\citep[see, e.g., Figure 4 of][for wind solutions]{kat94h}.
The velocity of the ``weak winds'' does not exceeds the escape velocity.
Winds are accelerated mainly by the peak of OPAL opacity due to
iron lines while they are sometimes accelerated by a small opacity
peak due to helium ionization for the relatively low metallicity region.
Here we regard that stripping effect works only for strong winds.
We have strong wind solutions (i.e., stripping effect)
for $M_{\rm WD} \gtrsim 0.6~M_\sun$
in the solar metallicity environment while for 
$M_{\rm WD} \gtrsim 1.0~M_\sun$ in a low metallicity region
of $Z=0.001$ (Population II), and further for 
$M_{\rm WD} \gtrsim 1.2~M_\sun$ in the much lower metallicity region
of [Fe/H]$\lesssim -2$.

Mass accretion onto the WD spins it up because angular
momentum is added to the WD \citep[e.g.,][]{lan00, pac91, pop91,
pie03a, pie03b, uen03}.
If the WD rotates rigidly, the mass of the WD can only slightly
exceed the Chandrasekhar mass limit of no rotation, $M_{\rm Ch}$.
More massive WDs seem to be possible if they rotate differentially
\citep[e.g.,][]{hac86}.
Based on simplified numerical calculations, \citet{yoo04} showed that
the WD can increase its mass beyond $M_{\rm Ch}$
when the accretion rate onto a WD is
high enough ($\dot{M}_{\rm WD} \gtrsim 10^{-7} M_\odot$~yr$^{-1}$)
and the WD is differentially rotating.
They showed that the gradient of rotational rate is kept around the  
critical value
for the dynamical shear instability, because the timescale for the
Eddington-Sweet meridional circulation is too long ($\sim 10^9$~yr) to
redistribute angular momentum in the WD core, and that
the differential rotation is strong enough for the WD mass to exceed
$M_{\rm Ch}$ significantly.

In the present study, we simply
assume that the WD is supported by differential rotation and its mass
can increase without carbon being fused
at the center as long as the mass accretion rate is higher than
$3\times 10^{-7}M_\odot$~yr$^{-1} (\equiv \dot{M}_{\rm b})$.
We have adopted the limit by considering the fact that
when $\dot M_{\rm WD} < \dot{M}_{\rm b}$
hydrogen shell-burning occurs intermittently and recurrent nova
outbursts eject a large part of the hydrogen-rich envelope
\citep[e.g.][]{hac01k}.
As a result, the net growth rate of the C+O core mass is reduced
significantly so that the timescale of the angular momentum deposition
on the WD would become comparable to the timescale for meridional  
circulation
to re-distribute the angular momentum.


\begin{figure*}
\epsscale{0.6}
\plotone{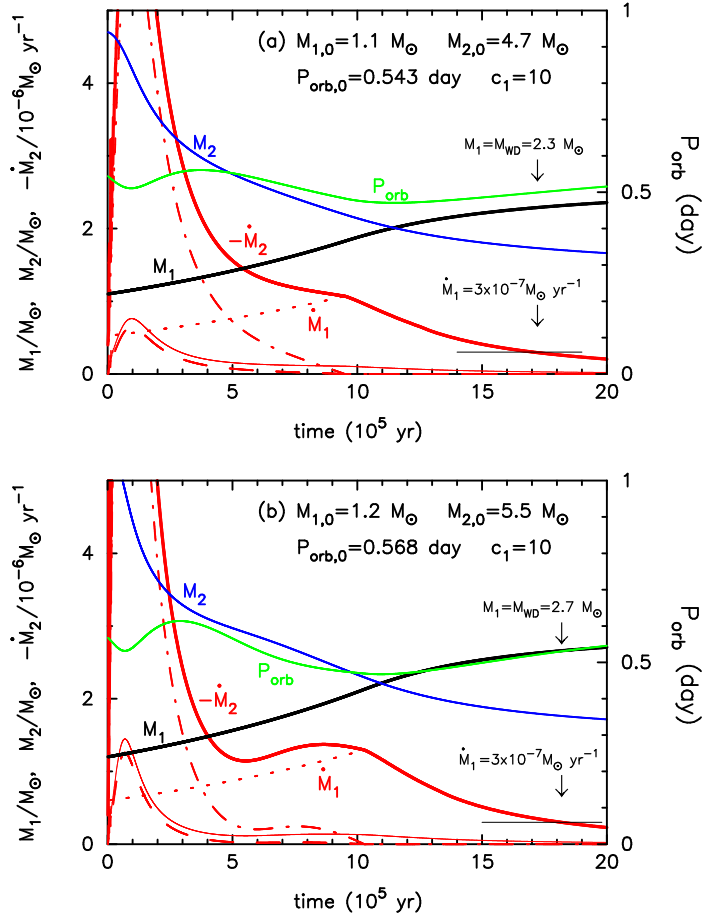}
\caption{
Binary evolutions of two WD+MS pairs: time starts
when the mass transfer begins from the MS to the WD.
(a) A WD of $M_{\rm WD,0} = 1.1~M_\sun$ and a MS star of
$M_{\rm MS, 0} = 4.7~M_\sun$ with the initial orbital
period of $P_{\rm orb,0}=0.543$~day (near ZAMS).
The red thick solid line: the mass-loss rate
from the MS companion, $-\dot M_2$.  The red thin solid line:
one tenth of the mass-loss rate, $-0.1\dot M_2$.
The other solid lines denote the companion mass ({\it blue}),
$M_2\equiv~M_{\rm MS}$, the orbital period ({\it green}),
$P_{\rm orb}$, and the WD (primary) mass ({\it black}),
$M_1\equiv~M_{\rm WD}$, each symbol
of which is attached to each line.  The red dash-dotted line:
stripping rate from the MS companion, $\dot M_{\rm strip}$. 
The long red dashed line: the wind mass-loss rate from the WD,
$\dot M_{\rm wind}$.  The red dotted line: the mass-increasing
rate of the WD, $\dot M_1\equiv\dot M_{\rm WD}$, which is merged
to the red solid line in the later time.  The mass-loss rate,
$-\dot M_2$, increases up to $6\times10^{-6}~M_\sun$~yr$^{-1}$
and quickly drops to $1\times10^{-6}~M_\sun$~yr$^{-1}$.  Then,
the optically thick wind stops and the mass-loss rate decreases
to $-\dot M_2=3\times10^{-7}~M_\sun$~yr$^{-1}$ at $t=1.7\times10^6$~yr.
At this time, the WD mass has increased to $M_1=M_{\rm WD}=2.3~M_\sun$.
(b) Same as (a) but for a pair of 1.2~$M_\sun$~WD and 5.5~$M_\sun$~MS
with the initial orbital period of $P_{\rm orb,0}=0.568$ day (near ZAMS).
The mass-loss rate increases up to
$-\dot M_2=1.5\times10^{-5}~M_\sun$~yr$^{-1}$
and then quickly drops to $1\times10^{-6}~M_\sun$~yr$^{-1}$.  Then,
the optically thick wind stops and the mass-loss rate decreases
to $-\dot M_2=3\times10^{-7}~M_\sun$~yr$^{-1}$ at $t=1.8\times10^6$~yr.  
At this time, the WD mass has increased to 
$M_1=M_{\rm WD}=2.7~M_\sun$.
\label{gkper_evol_binary_m470_wd11_m550_wd12_combine_c1_10}}
\end{figure*}


\begin{figure*}
\epsscale{0.6}
\plotone{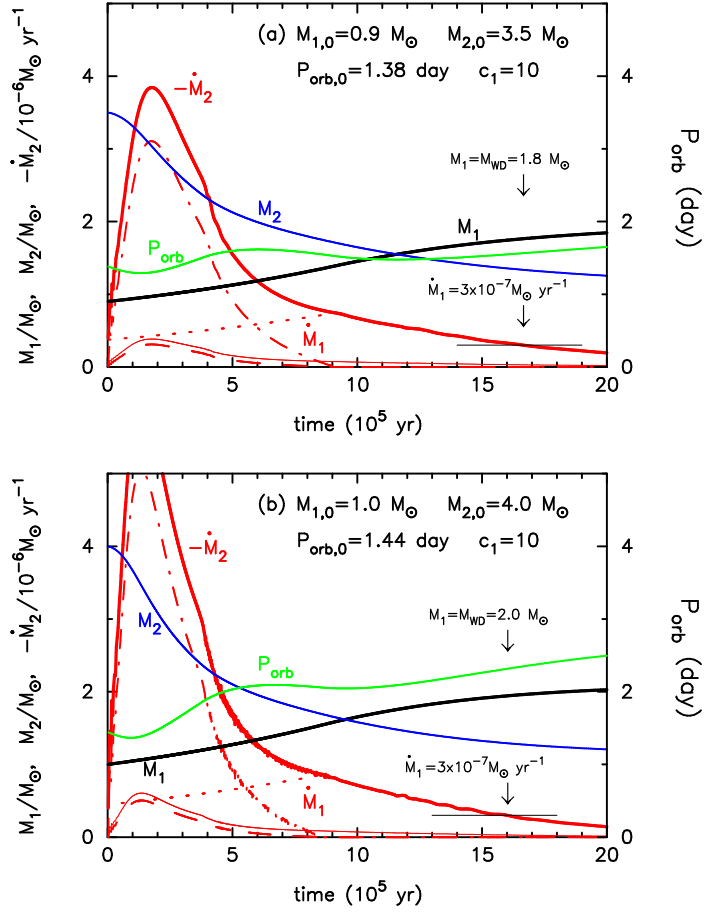}
\caption{
Same as Figure\ref{gkper_evol_binary_m470_wd11_m550_wd12_combine_c1_10},
but for the companion models near the end of central hydrogen-burning.
(a) A WD of $M_{\rm WD,0} = 0.9~M_\sun$ and a MS of
$M_{\rm MS, 0} = 3.5~M_\sun$ with the initial orbital
period of $P_{\rm orb,0}=1.38$ day.
The mass-loss rate from the companion decreases
to $-\dot M_2=3\times10^{-7}~M_\sun$~yr$^{-1}$ at $t=1.67\times10^6$~yr.
At this time, the WD mass has increased to 1.8~$M_\sun$.
(b) A WD of $M_{\rm WD,0} = 1.0~M_\sun$
and a MS of $M_{\rm MS, 0} = 4.0~M_\sun$
with the initial orbital period of $P_{\rm orb,0}=1.44$~day.
The mass-loss rate decreases
to $-\dot M_2=3\times10^{-7}~M_\sun$~yr$^{-1}$ at $t=1.59\times10^6$~yr.  
At this time, the WD mass has increased to 2.0~$M_\sun$.
\label{gkper_evol_binary_m350_wd09_m400_wd10_combine_c1_10}}
\end{figure*}

\section{Numerical Results}
\label{numerical_results}
\subsection{Evolutions of WD+MS Systems}
We have investigated evolutions of binaries which initially consist
of a WD (primary) and a zero-age main-sequence (ZAMS) star,
by calculating the evolutions of the secondary stars
using a Henyey-type code \citep[e.g.,][]{sai88}.
We have adopted a standard chemical composition 
of $(X,Z)=(0.70,0.02)$ and used OPAL opacity tables \citep{opal95}.
The mixing length is assumed to be 1.5 times the pressure
scale height.

As the star evolves, the radius increases and eventually fills
its critical Roche lobe. Then, a mass transfer from the secondary to the
primary WD begins.  We have determined the rate of mass loss 
from the secondary, $\dot M_2$, at each stage by requiring
for the stellar radius to be equal to the mean radius of the Roche lobe.
If $|\dot M_2|$ is smaller than the critical rate
$\dot M_{\rm cr}$ given in Equation~(\ref{steady_accretion_wind}),
all the matter from the secondary 
will be accreted onto the WD; i.e., $\dot M_1 = |\dot M_2|$.
When the mass-accretion rate exceeds $\dot M_{\rm cr}$,  
an optically thick wind blows from the WD.
It causes mass-stripping from the secondary star as 
discussed in the previous section.
Then, the systemic mass loss at a rate of
$\dot M_{\rm wind} + \dot M_{\rm strip}$ occurs, which 
affects the evolution of the Roche lobe radius and hence $\dot M_2$.
During the evolution, $\dot M_1$ is given as
$\min(\dot M_{\rm cr}, |\dot M_2|) $
\citep[see][for details]{hkn96, hkn99, hkn08a}.

In the previous work \citep{hkn08a}, we stopped the binary evolution
at the epoch when the WD mass reached $1.38~M_\sun$,
because we regarded that a WD explodes as an SN~Ia at
$M_{\rm WD}=1.38~M_\sun$ \citep{nom82}.
In the present work, however, we further follow
the binary evolution until the mass 
accretion rate of the primary WD decreases to
$\dot M_1=\dot M_{\rm b}=3\times 10^{-7}~M_\sun$~yr$^{-1}$,
because the WD is expected to be supported by a strong differential
rotation until then.

Although Equation (\ref{steady_accretion_wind})
is based on the assumption of spherical symmetry \citep{nom07},
we consider it approximately applicable even  
for $M_{\rm WD}>1.38~M_\sun$ for the following reasons. 
Equation (\ref{steady_accretion_wind}) is closely related
to the local Eddington luminosity (and nuclear energy conversion rate
per unit mass for hydrogen mass fraction, $X=0.70$), which 
is mainly determined by the local opacity (peak) at 
a part with temperature of $\sim 2\times10^5$~K in the envelope.
Since this layer is far from the C+O core surface (at the hydrogen shell
burning region), the local Eddington luminosity
should be insensitive to a rapid rotation of the C+O core.


\begin{figure*}
\epsscale{0.6}
\plotone{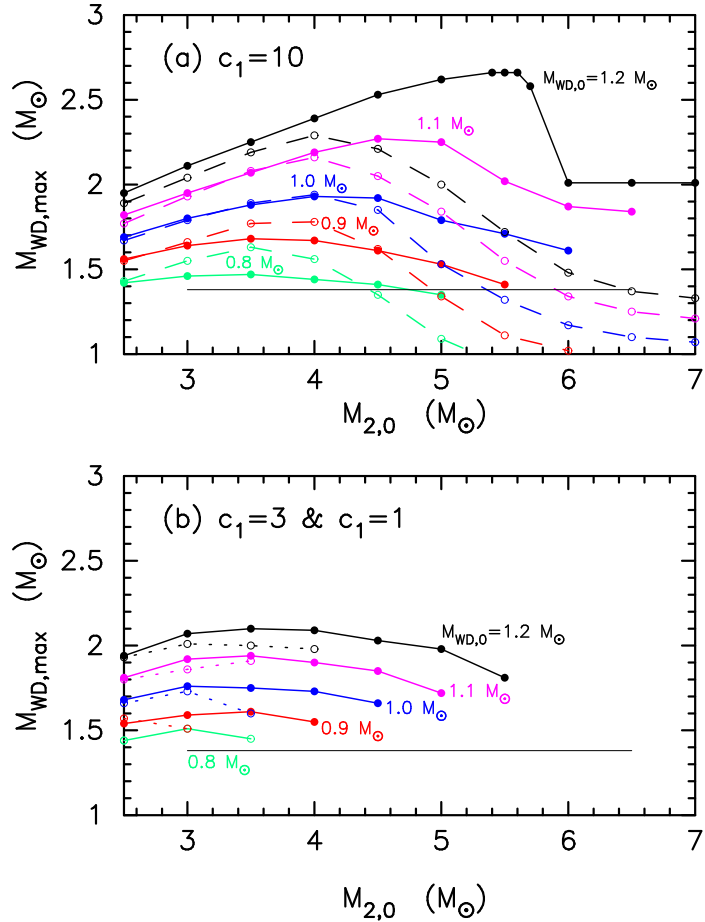}
\caption{
The maximum WD mass against the companion mass for the various 
initial WD masses and orbital periods. (a) $c_1=10$.
The solid lines with small filled circle:
the companion is slightly evolved off from ZAMS but still
very close to ZAMS ($P_{\rm orb,0}\sim 0.5$~day) when the companion
starts mass-transfer.
The dashed lines with small open circle: the companion is close to
the end of central hydrogen-burning  ($P_{\rm orb,0}\sim 1.5$~day)
when the companion starts mass-transfer.
The horizontal thin solid line denotes $M_{\rm WD}=1.38~M_\sun$.
(b) Same as those in (a) but for $c_1=3$ (solid lines)
and $c_1=1$ (dashed lines), where the companion fills its Roche lobe
near ZAMS, i.e., $P_{\rm orb,0}\sim 0.5$~day.  See text for more details.
\label{maximum_wd_mass_c10}}
\end{figure*}

We have followed binary evolutions for various sets of the initial
parameters $(M_{\rm WD,0}, M_{2,0}, P_{\rm orb,0})$
and obtained maximum WD masses.
In the first example, we set $c_1=10$ and assume
$M_{\rm WD,0}=1.1~M_\sun$, $M_{2,0}=4.7~M_\sun$, and
$P_{\rm orb,0}=0.54$~day (the companion fills its Roche lobe after
it slightly evolved off from ZAMS).  The result is shown in
Figure \ref{gkper_evol_binary_m470_wd11_m550_wd12_combine_c1_10}a.
The mass-loss rate of the companion quickly increases up to
$-\dot M_2=8\times10^{-6}M_\sun$~yr$^{-1}$ and then drops 
to $-\dot M_2=1\times10^{-6}M_\sun$~yr$^{-1}$ at 
$t=10^6$~yr ($-\dot M_2$ is also plotted in the one tenth scale
in the same figure  by a red thin solid line).
In the early phase of mass transfer, we have 
$-0.1\dot M_2 \approx\dot M_{\rm wind}$ because 
$-\dot M_2 \approx\dot M_{\rm strip}$ and $c_1=10$.
As mass transfer rate decreases,
the optically thick wind stops at $t\sim1\times10^6$~yr,
and the accretion rate $\dot M_1$ decreases 
to $3\times10^{-7}~M_\sun$~yr$^{-1}$ at $t=1.7\times10^6$~yr.
The WD mass has increased to 2.3$~M_\sun$ at this epoch.
(Even if we adopt $\dot M_{\rm b}=1$, 2, 3, and
$4\times10^{-7}~M_\sun$~yr$^{-1}$, the maximum WD mass does not change
so much, i.e., 2.45, 2.36, 2.29, and 2.22$~M_\sun$, respectively.
The dependence of the WD mass on the value of 
$\dot M_{\rm b}$ is very weak so that the adopted value itself is
not essential for the results.)

In the second example, we assume a more massive binary with
$c_1=10$, $M_{\rm WD,0}=1.2~M_\sun$, $M_{2,0}=5.5~M_\sun$, and
$P_{\rm orb,0}=0.57$~day (the companion fills its Roche lobe near ZAMS
as in the first case).  The result is shown in
Figure \ref{gkper_evol_binary_m470_wd11_m550_wd12_combine_c1_10}b.
The mass-loss rate of the companion increases up to
$-\dot M_2=1.4\times10^{-5}M_\sun$~yr$^{-1}$ and then quickly
decreases.   The optically thick wind stops at almost
the same time as in Figure
\ref{gkper_evol_binary_m470_wd11_m550_wd12_combine_c1_10}a,
$t=1\times10^6$~yr, and afterward
the mass transfer rate gradually decreased to 
$\dot M_1=\dot M_{\rm b}=3\times10^{-7}~M_\sun$~yr$^{-1}$
at $t=1.8\times10^6$~yr.
The WD mass has increased to 2.7 $M_\sun$ at this epoch.

The third and forth examples are for the binaries in which the companion
fills its Roche lobe in a later phase, i.e.,
near the end of central hydrogen burning
(Figure \ref{gkper_evol_binary_m350_wd09_m400_wd10_combine_c1_10}).
The third example is for $c_1=10$, $M_{\rm WD,0}=0.9~M_\sun$,
$M_{2,0}=3.5~M_\sun$, and $P_{\rm orb,0}=1.38$~day.  The result is shown
in Figure \ref{gkper_evol_binary_m350_wd09_m400_wd10_combine_c1_10}a.
The mass-loss rate of the companion increases up to
$-\dot M_2=3.8\times10^{-6}M_\sun$~yr$^{-1}$ and then quickly
decreases.   The optically thick wind stops at 
$t=9\times10^5$~yr, and afterward
the mass transfer rate is gradually decreasing to 
$\dot M_{\rm b}=3\times10^{-7}~M_\sun$~yr$^{-1}$ at $t=1.7\times10^6$~yr.
The WD mass has increased to 1.8$~M_\sun$ at this epoch.

In the fourth example, we have adopted $c_1=10$, $M_{\rm WD,0}=1.0~M_\sun$,
$M_{2,0}=4.0~M_\sun$, and $P_{\rm orb,0}=1.44$~day. 
The result is shown in
Figure \ref{gkper_evol_binary_m350_wd09_m400_wd10_combine_c1_10}b.
The mass-loss rate of the companion increases up to
$-\dot M_2=6\times10^{-6}M_\sun$~yr$^{-1}$ and then quickly
decreases.   The optically thick wind stops at 
$t=8\times10^5$~yr, and afterward
the mass transfer rate is gradually decreasing to 
$\dot M_{\rm b}=3\times10^{-7}~M_\sun$~yr$^{-1}$ at $t=1.6\times10^6$~yr.
The WD mass has increased to 2.0$~M_\sun$ at this epoch.

In all the models, the mass-loss rate from the secondary,
$-\dot M_2$, increases to $\gtrsim3\times 10^{-6}~M_\sun$~yr$^{-1}$.
If all the hydrogen-rich matter were converted into helium and then
into carbon-oxygen and therefore accreted at the same rate as $-\dot M_2$, 
carbon burning would ignite off-center and the C+O core
would been converted into O+Ne+Mg core \citep[e.g.,][]{kaw88}
so that the WD would not explode as an SN~Ia.
However, the net accretion rate to the WD is given as 
$|\dot M_2|-|\dot M_{\rm strip}|-|\dot M_{\rm wind}|$
($=\dot M_1 \equiv \dot M_{\rm WD}=$~the dotted lines in Figures
\ref{gkper_evol_binary_m470_wd11_m550_wd12_combine_c1_10}
and \ref{gkper_evol_binary_m350_wd09_m400_wd10_combine_c1_10}),
not exceeding $3\times 10^{-6} M_\sun$~yr$^{-1}$.  So
we expect no off-center burning of carbon during
the accretion phase and that the C+O core grows until carbon
ignites at the center.   

Our numerical results are summarized in Figure \ref{maximum_wd_mass_c10}.
This figure shows the maximum WD mass $M_{\rm WD, max}$ obtained
in the binary evolution against the initial companion mass $M_{2,0}$
for various sets of parameters, $M_{\rm WD,0}$, $P_{\rm orb,0}$, and $c_1$.
Figure \ref{maximum_wd_mass_c10}a shows examples for the parameter of
$c_1=10$.  We set a step of 0.5$~M_\sun$ for the initial companion mass
(except for a few models).  The rightmost ends of these solid lines
correspond to the model beyond which the mass transfer rate of
$-\dot M_2$ quickly increases to $10^{-4}$--$10^{-3}~M_\sun$~yr$^{-1}$
and thus a common envelope is formed.
We regard that two stars are merging into one.
Solid lines are for the case that the companion fills
its Roche lobe near ZAMS, i.e., $P_{\rm orb,0} \sim 0.5$ day. 
The WD mass reaches $M_{\rm WD,max} \approx 
2.7$, 2.3, 1.9, 1.7, and 1.5$~M_\sun$, for the initial WD mass
of $M_{\rm WD,0}= 1.2$, 1.1, 1.0, 0.9, and 0.8$~M_\sun$, respectively.
The dashed lines show the case that
the companion fills its Roche lobe near the end of central
hydrogen burning, i.e., $P_{\rm orb,0}\sim 1.5$~day.  In this case,
$M_{\rm WD,max} \approx 2.3$, 2.2, 2.0, 1.8, and 1.6$~M_\sun$,
for $M_{\rm WD,0}= 1.2$, 1.1, 1.0, 0.9, and 0.8$~M_\sun$, respectively.
For the latter case (dashed lines) $M_{\rm WD,max}$ smaller than
that for the former case (solid lines) even for the same $c_1$,
$M_{\rm WD,0}$, and $M_{2,0}$.  This is because the more evolved
secondary has a larger mass transfer rate, $-\dot M_2$, and thus
the duration of a high mass-accretion phase, in which period
the WD grows quickly, is shorter than that of the near ZAMS case.

Figure \ref{maximum_wd_mass_c10}b shows examples for the parameters
of $c_1=3$ (solid lines) and $c_1=1$ (dotted lines), in which
the companion fills its Roche lobe
near ZAMS, i.e., $P_{\rm orb,0}\sim 0.5$~day.
The obtained maximum WD mass is smaller than that for $c_1=10$.
In our previous work \citep{hkn08a}, based on a simplified model
of mass-transfer, we showed that more massive companions
can avoid merging for larger $c_1$-parameters.  In the present work,
we adopted a more realistic companion model which results in more
accurate mass-transfer rates.
Comparing Figure \ref{maximum_wd_mass_c10}a
with Figure \ref{maximum_wd_mass_c10}b, we confirm that more massive
companion avoids merging for a larger $c_1$-parameter.

The maximum WD mass depends strongly on $c_1$ and $M_{\rm WD,0}$.
To summarize, the maximum WD masses are
$M_{\rm WD, max}=1.9$, 1.9, and $2.3~M_\sun$ for $c_1=1$, 3, and 10,
if the binary evolution starts from $M_{\rm WD,0}=1.1~M_\sun$.
If the binary evolution starts from a more massive initial WD mass
of $M_{\rm WD,0}=1.2~M_\sun$, a larger maximum WD mass is resulted as:
$M_{\rm WD, max}=2.0$, 2.1, and $2.7~M_\sun$ for $c_1=1$, 3, and 10,
respectively.


\begin{figure*}
\epsscale{0.8}
\plotone{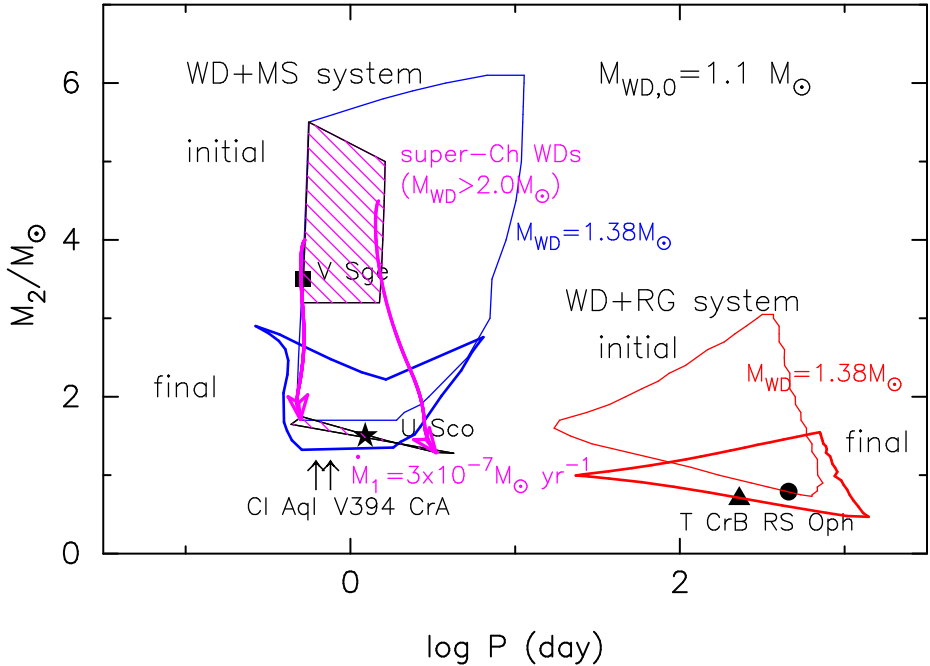}
\caption{
The regions that produce super-Chandrasekhar mass ($M_{\rm WD} 
> 2.0 ~M_\sun$) SNe~Ia are plotted in the $\log P - M_2$
(orbital period -- secondary mass) plane (hatched by magenta lines).
Here we assume the metallicity of $Z=0.02$ and
the initial WD mass of $M_{\rm WD, 0}= 1.1 ~M_\sun$.
The initial system inside the hatched region (above)
increases its WD mass up to a super-Chandrasekhar mass of
$M_{\rm WD} > 2.0~M_\sun$ and then reaches the hatched
region (below) where the mass-transfer rate drops
below $\dot M_b =3 \times 10^{-7} M_\sun$~yr$^{-1}$.
Evolutionary paths of two typical cases ($4.0 ~M_\sun$ companion
star near ZAMS and $4.5 ~M_\sun$ companion near the end of
central hydrogen burning) are traced by a thick
magenta line with an open arrow.  We also added the previous results
taken from \citet{hkn08a,hkn08b,hkn99},
where we assumed that a non-rotating C+O WD explodes
at $M_{\rm WD}=1.38~M_\sun$.
The initial system inside the region encircled by a thin solid line
(labeled ``initial'') increases its WD mass up to the critical
mass ($M_{\rm WD}= 1.38 M_\sun$) and explodes in
the regions encircled by a thick solid line
(labeled ``final'') for the white dwarf and
main-sequence star (WD + MS) system ({\it left}) and the white dwarf
and red giant (WD + RG) system ({\it right}).
Currently known positions of Galactic recurrent novae
and supersoft X-ray sources are indicated by a star mark ($\star$)
for U Sco \citep[e.g.,][]{hkkm00}, a triangle for T CrB
\citep{bel98}, a filled circle for RS Oph
\citep{bra09}, a square for V Sge \citep{hac03kc}.
Arrows indicate V394 CrA and CI Aql, because their companion masses
are not known.  
\label{zregevl10_strip_ms_rg_wd11}}
\end{figure*}

\subsection{Initial-final States of Binary Evolutions}
We plot, in Figure \ref{zregevl10_strip_ms_rg_wd11},
the initial parameter region
that produces super-Chandrasekhar mass ($M_{\rm WD} 
> 2.0 ~M_\sun$) SNe~Ia in the $\log P - M_2$
(orbital period -- secondary mass) diagram (hatched by magenta lines).
Here we assume the metallicity of $Z=0.02$ and
the initial white dwarf mass of $M_{\rm WD, 0}= 1.1 ~M_\sun$.
A binary system initially inside the hatched region (above)
increases its WD mass up to a super-Chandrasekhar mass of
$M_{\rm WD} > 2.0~M_\sun$.  The companion mass decreases with time,
so that its position moves downward.  When it reaches the hatched
region (below), the mass-transfer rate drops
below $\dot M_b =3 \times 10^{-7} M_\sun$~yr$^{-1}$ and
the Eddington-Sweet meridional circulation would redistribute 
angular momentum in the WD to trigger central carbon burning.
Evolutionary paths of two typical cases of $M_{2,0}=4.0 ~M_\sun$ companion
star near ZAMS ($P_0=0.53$ day) and $M_{2,0}=4.5 ~M_\sun$ companion near
the end of central hydrogen burning ($P_0=1.48$ day) are followed by a thick
magenta line with an arrow.  Our calculated final states fall into
a very narrow strip because all their evolutions stop at the same
mass transfer rates of $\dot M_b=3 \times 10^{-7} M_\sun$~yr$^{-1}$.

For comparison, we also added our previous results
taken from \citet{hkn08a,hkn08b,hkn99},
where we assumed that a non-rotating C+O WD explodes
at $M_{\rm WD}=1.38~M_\sun$ \citep{nom82}.  The initial binary system
inside the region encircled by a thin solid line
(labeled ``initial'') increases its WD mass up to the critical
mass ($M_{\rm WD}= 1.38 M_\sun$) and explodes in
the regions encircled by a thick solid line
(labeled ``final'') for the white dwarf and
main-sequence star (WD + MS) system ({\it left}) and the white dwarf
and red giant (WD + RG) system ({\it right}).
We have found no super-Chandrasekhar mass SN~Ia ($M_{\rm WD} 
> 2.0 ~M_\sun$) region for the WD+RG systems mainly because
mass of the companion is too small to supply much matter to the WD. 
Currently known positions of some Galactic recurrent novae
and supersoft X-ray sources are indicated by a star mark ($\star$)
for U Sco \citep[e.g.,][]{hkkm00}, a triangle for T CrB
\citep{bel98}, a filled circle for RS Oph
\citep{bra09}, a square for V Sge \citep{hac03kc},
and by arrows for the two recurrent novae, V394 CrA and
CI Aql, because mass of a companion is not yet
accurately obtained.  Two subclasses of the recurrent novae, the
U Sco type and the RS Oph type, correspond to the WD + MS
channel and the WD + RG channel of SNe~Ia, respectively.

\citet{hac01k} summarized the mass transfer rates of Galactic
recurrent novae in their Table 2, in which only U Sco has a large mass
transfer rate enough to support differential rotation, i.e.,
the mass transfer rate is $\sim 2.5 \times 10^{-7} M_\sun$~yr$^{-1}$
and the net growth rate of C+O core is
$\sim 1.0 \times 10^{-7} M_\sun$~yr$^{-1}$.  Therefore, U Sco may
have a super-Chandrasekhar mass WD.  \citet{tho01} observationally
estimated the WD mass of U Sco to be $M_{\rm WD}= 1.55 \pm 0.24
~M_\sun$ in the 1999 outburst.  The systematic error is, however,
too large to draw a definite conclusion for super-Chandrasekhar mass,
so we strongly encourage more accurate observations to determine
the WD mass in U Sco.


\begin{figure*}
\epsscale{0.7}
\plotone{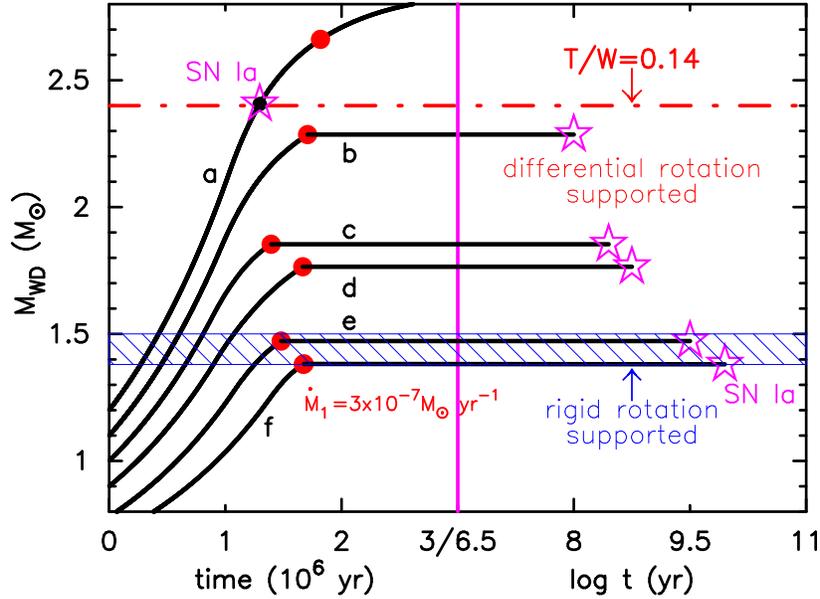}
\caption{
Schematic illustration for evolutions of the WD mass against time.
Time starts when the secondary fills its Roche lobe and it is switched from
linear to logarithmic at $t\approx 3 \times 10^6$~yr (a vertical
magenta line).  The evolutional lines in the left (part) to the magenta
line are taken from our numerical results
while those in the right (part) are just schematic illustrations. 
From top to bottom:
(a) $M_{\rm WD,0}=1.2~M_\sun$, $M_{2,0}=5.5~M_\sun$, $P_0=0.57$~day, 
$c_1=10$.  Same model as the model in Figure 
\ref{gkper_evol_binary_m470_wd11_m550_wd12_combine_c1_10}b.
In this model, $T/|W|$ exceeds 0.14 before
the mass accretion rate becomes lower than
$\dot M_{\rm b}= 3\times10^{-7}~M_\sun$~yr$^{-1}$.
We expect an SN~Ia explosion at the position of magenta star mark
($M_{\rm WD}\approx 2.4~M_\sun$) soon after a secular instability sets in.
(b) $M_{\rm WD,0}=1.1~M_\sun$, $M_{2,0}=4.7~M_\sun$, $P_0=0.54$~day, 
$c_1=10$.  Same model as the model in Figure 
\ref{gkper_evol_binary_m470_wd11_m550_wd12_combine_c1_10}a,
where $T/|W|$ does not exceed 0.14 before
$\dot M_1 < \dot M_{\rm b}= 3\times10^{-7}~M_\sun$~yr$^{-1}$.
Loss or redistribution of angular momentum leads to an increase in
the central density until carbon is ignited to induce SN Ia explosion
($M_{\rm WD}\approx 2.3~M_\sun$).  The delay time depends
on the timescales of angular momentum loss or redistribution.
It may take a time of order $10^8$--$10^9$~yr
if angular momentum redistribution occurs due to Eddington-Sweet
circulation \citep{yoo04}.
(c) $M_{\rm WD,0}=1.0~M_\sun$, $M_{2,0}=4.5~M_\sun$, $P_0=1.46$~day, 
$c_1=10$.  The WD mass reaches $M_{\rm WD}=1.85~M_\sun$, which is
differential-rotation-supported.
(d) $M_{\rm WD,0}=0.9~M_\sun$, $M_{2,0}=3.5~M_\sun$, $P_0=1.38$~day, 
$c_1=10$.  The WD mass reaches $M_{\rm WD}=1.76~M_\sun$, which is
differential-rotation-supported.
(e) $M_{\rm WD,0}=0.8~M_\sun$, $M_{2,0}=4.9~M_\sun$, $P_0=0.52$~day, 
$c_1=10$.  The WD mass reaches $M_{\rm WD}=1.47~M_\sun$, which can be
supported by rigid rotation \citep{hac86, uen03}.
Therefore, it may take a longer time
than the case of differentially-rotating-supported WDs.
(f) $M_{\rm WD,0}=0.7~M_\sun$, $M_{2,0}=3.0~M_\sun$, $P_0=0.47$~day, 
$c_1=5$.    The WD mass reaches $M_{\rm WD}=1.38~M_\sun$, which can 
also be supported by rigid rotation.  
\label{spindown_evol_ms_wd_c1_10}}
\end{figure*}

\subsection{Stability of Mass-accreting, Differentially Rotating WDs} 
In the present paper, we simply assume that the mass-accreting,
differentially rotating WDs are
stable and can accrete mass from the companion as long as its
mass accretion rate is larger than
$M_{\rm b}=3\times10^{-7}~M_\sun$~yr$^{-1}$.
This assumption is based on the results in \citet{yoo04}.
However, they did not incorporate the possibility of baroclinic
instability in the WD mainly because equi-density and equi-pressure
lines are essentially parallel in the WD due to high degeneracy.
If the baroclinic instability could occur \citep{fuj93},
angular momentum is effectively transported in a shorter timescale
than the accretion timescale, so that a rigid rotation is realized
\citep{sai04}.  If this is the case, suggested super-Chandrasekhar
mass WDs would not be realized in the SD scenario.

Recently, \citet{pir08} concluded, based on Fujimoto's (1993)
baroclinic instability criterion, that baroclinic instabilities 
bring the WD very close to solid-body rotation during accretion phase,
which makes it difficult to grow a WD mass substantially past
the Chandrasekhar mass.  However, we should notice that Fujimoto's
criterion on baroclinic instability is just a necessary condition but
not a sufficient condition.  In this sense, Piro's (2008) result
is not always correct.  We must keep in mind 
that he showed only a possibility for baroclinic instability
on mass-accreting WDs but not an occurrence of baroclinic
instability.  On the other hand, the Richardson criterion on
the Kelvin-Helmholtz instability is also a necessary condition
but has ever been tested for a variety of basic (plane-parallel) flows
as a sufficient condition for the instability 
\citep[see, e.g.,][]{fuj93}.  Yoon \& Langer's (2005) stability
criterion was based on the Richardson criterion of the
Kelvin-Helmholtz instability.  Therefore, we expect that 
the true mixing instability for redistribution of angular momentum
occurs somewhere between
the Richardson criterion and Fujimoto's (1993) criterion.

If we assume Yoon \& Langer's (2005) stability criterion 
(the Richardson criterion) and
extrapolate their formula on $T/|W|$ (their Equation [29])
to higher masses, where $T$ and $W$ are the rotational and gravitational
energies of the WD, respectively, 
angular momentum redistribution could occur 
when $T/|W|$ exceeds 0.14 and a secular instability sets in.
This instability occurs at $M_{\rm WD} \approx 2.4~M_\sun$
\citep[see also][]{hac86}.
At this time, the WD would be too massive to keep
hydrostatic equilibrium and contracts on a much shorter timescale
due to angular momentum transport.
As a result of such a rapid compression, the WD eventually
ignites explosive carbon burning in the central region to give rise to
an SN~Ia explosion.  If super-Chandrasekhar mass SNe~Ia occur
in this way, the WD mass can be as massive as up to $2.4~M_\sun$.
To confirm this prediction, accurate estimates of
the ejecta masses, in which asphericity is taken into account,
would be important.

\subsection{Evolutions in Three Mass Ranges of Exploding WDs}
If we apply the above discussion to the super-Chandrasekhar mass WDs,
we may further discuss the fate of the WDs.
There are three characteristic mass ranges of exploding WDs depending
on the secular instability and rotation law:
(1) $M_{\rm WD}>2.4~M_\sun$ (a secular instability at $T/|W|=0.14$),
(2) $M_{\rm WD}=1.5$--2.4$~M_\sun$ (differential rotation),
and (3) $M_{\rm WD}=1.38$--1.5$~M_\sun$ (rigid rotation).

\subsubsection{SNe Ia triggered by Secular Instability:
SN~2007if/SN~2009dc-like}
For very massive WDs, a secular instability sets in at $T/|W|=0.14$
and the WD explodes as an SN~Ia when it reaches $2.4~M_\sun$.
For example, the model of $M_{\rm WD,0}=1.2~M_\sun$,
$M_{2,0}=5.5~M_\sun$, $P_0=0.57$~day, and $c_1=10$ 
(Figure \ref{gkper_evol_binary_m470_wd11_m550_wd12_combine_c1_10}b)
reaches $T/|W|=0.14$ at $t=1.28\times 10^6$~yr.  This evolution model is
also shown in Figure \ref{spindown_evol_ms_wd_c1_10} by 
the uppermost line a (``a'' is attached to the line).
Therefore, we expect an SN~Ia explosion to occur at the 
magenta star mark because the timescale of the secular instability is
much shorter than the mass-accreting timescale.
This kind of binary evolution corresponds to the
very luminous super-Chandrasekhar mass SNe~Ia such as
SN~2007if and SN~2009dc.  It is interesting that these brightest
super-Chandrasekhar mass SNe~Ia are the youngest among our models
(younger than a few hundred Myr).

\subsubsection{SNe Ia triggered by Loss or Redistribution of
Angular Momentum: SN~2003fg/SN~2006gz-like}
For $M_{\rm WD}=1.5$--2.4$~M_\sun$, $T/|W|$ does not exceed 0.14.
For example, the model for the initial parameters of
$M_{\rm WD,0}=1.1~M_\sun$, $M_{2,0}=4.7~M_\sun$, $P_0=0.54$~day, 
and $c_1=10$ (Figure 
\ref{gkper_evol_binary_m470_wd11_m550_wd12_combine_c1_10}a)
reaches $M_{\rm WD}=2.3~M_\sun$ before the accretion rate drops to
$\dot M_1 < \dot M_{\rm b}= 3\times10^{-7}~M_\sun$~yr$^{-1}$
at the red filled circle in Figure \ref{spindown_evol_ms_wd_c1_10}
(line b).
We also show two other examples. One is for the initial parameters
of $M_{\rm WD,0}=1.0~M_\sun$, $M_{2,0}=4.5~M_\sun$, $P_0=1.46$~day, and
$c_1=10$.  In this case, the WD mass reaches $M_{\rm WD}=1.85~M_\sun$,
which is also differential-rotation-supported (line c in
Figure \ref{spindown_evol_ms_wd_c1_10}).
The other is for
$M_{\rm WD,0}=0.9~M_\sun$, $M_{2,0}=3.5~M_\sun$, $P_0=1.38$~day, and 
$c_1=10$.  The WD mass reaches $M_{\rm WD}=1.76~M_\sun$, which is
again differential-rotation-supported
(line d in Figure \ref{spindown_evol_ms_wd_c1_10}).

  However, the WD does not explode soon after the WD mass
has reached the maximum mass (1.5--$2.3~M_\sun$),
because the central density is low so that
central carbon is not ignited yet.  After some time,
loss or redistribution of angular momentum would lead to an increase
in the central density of the WD.  As a result, carbon is ignited to
induce an SN~Ia.  This waiting time depends
entirely on the timescale of angular momentum loss or redistribution.
It is an order of $10^8$--$10^9$~yr
if angular momentum redistribution occurs due to the Eddington-Sweet
circulation \citep{yoo04}.  If direct loss of angular momentum
occurs in a much shorter timescale (due to magneto-dipole radiation
or so), it could be much shorter than $\sim10^8$~yr.
In any case, it is shorter than
$10^8$--$10^9$~yr.  Therefore we suggest that these relatively young
SNe~Ia correspond to the ``prompt'' component.  
We regard that this kind of binary evolution corresponds to 
super-Chandrasekhar mass SNe~Ia such as SN~2003fg/SN~2006gz,
not so extremely luminous as SN~2007if/SN~2009dc.

\subsubsection{SNe Ia triggered by Loss of Angular Momentum:
normal SNe~Ia}
For $M_{\rm WD} < 1.5~M_\sun$, WDs could be rigidly rotating.
For example, 
line e in
Figure \ref{spindown_evol_ms_wd_c1_10} shows the case of
$M_{\rm WD,0}=0.8~M_\sun$, $M_{2,0}=4.9~M_\sun$, $P_0=0.52$~day, and
$c_1=10$.  The WD mass reaches $M_{\rm WD}=1.47~M_\sun$, which can be
supported by rigid rotation \citep[e.g.,][]{hac86, uen03}.
Also, line f in
Figure \ref{spindown_evol_ms_wd_c1_10} depicts the case of initial
parameters of $M_{\rm WD,0}=0.7~M_\sun$, $M_{2,0}=3.0~M_\sun$,
$P_0=0.47$~day, and $c_1=5$.  
The WD mass reaches $M_{\rm WD}=1.38~M_\sun$
at the red filled circle.  The WD can 
also be supported by rigid rotation.

The WD will eventually explode as an SN~Ia after the WD spins
down and the central density increases high enough to ignite carbon.
This spinning-down timescale depends highly on the subtraction mechanism 
of angular momentum from the WD such as
magneto-dipole radiation in a strongly magnetized WD.
Recently, \citet{ilk11} reexamined the timescale
of spin-down due to r-mode gravitational radiation and showed
that r-mode instability is not significant in spinning down WDs
\citep[see also][]{lin99, yoo05}. 
Instead, they suggested that magneto-dipole radiation leads to
spinning-down in a typical timescale of $< 10^9$~yr in WDs as massive
as $M_{\rm WD} \gtrsim 1.6~M_\sun$ but in a typical
timescale of $> 10^9$~yr in $M_{\rm WD}=1.4$--$1.5~M_\sun$.
If it is the case, these rigidly-rotating WDs of
$M_{\rm WD}=1.38$--$1.5~M_\sun$ may correspond to the ``tardy'' component,
suggesting that the companion has evolved off to a white dwarf
when the WD explodes as an SN~Ia \citep{jus11, dis11}.  

There is an example of WDs that had once been extremely spun-up
but now are very slowly spinning down. 
Recently a massive, fast spinning WD with a spin period of
$P_{\rm spin}=13.2$~s and a mass of $M_{\rm WD}=1.28\pm 0.05~M_\sun$
was found in a binary system HD~49798/RX~J0648.0$-$4418
($P_{\rm orb}=1.55$~day), which is consisting of a hot subdwarf
and a WD \citep[e.g.,][]{mer11}.
This spin period is only a factor of two longer than the critical
spin period\footnote{
The total angular momentum and rotational energy are
$J=0.749 \times 10^{50}$~erg~s and $T=0.315 \times 10^{50}$~erg,
respectively, at the critical rotation from Table 4 of \citet{hac86}.
Then we have $\Omega_{\rm cr} = 2 T / J = 0.841$~rad~s$^{-1}$, so
the critical spin period is $P_{\rm spin,cr}=2\pi / \Omega_{\rm cr}
=7.47$~s for a rigidly
rotating 1.28~$M_\sun$ WD.}.
The spin-down rate is too small to be detected, i.e.,
$\dot P_{\rm spin} < 9 \times 10^{-13}$~s~s$^{-1}$.
\citet{mer11} estimated a very weak magnetic field of
$B < 1$~kG at the WD surface.  In such a low magnetic field,
the spin-down rate due to magneto-dipole radiation
should be so small that we expect a very long
decay time ($> 10^9$~yr) of the rapidly rotating WD
\citep[see, e.g., Equation 10 of][]{ilk11}.

\section{Possible Evolutionary Paths to Massive C+O WD Binaries}
In the previous section, we present candidate binaries that have
an evolutionary path toward a super-Chandrasekhar mass progenitor.
Such binaries should consist of a very massive C+O WD ($M_{\rm WD,0} 
\sim 1.1$--1.2$~M_\sun$) and an intermediate-mass
($M_{2,0} \sim$ 4--5$~M_\sun$) companion star
with an orbital period of $P_{\rm orb,0} \sim 0.5$--1.5 days.
These massive C+O WDs, however, have not been expected mainly
because dust-driven wind mass-loss prevents the C+O WDs
from growing in their mass in the AGB phase \citep[e.g.,][for
a recent review of AGB stars]{her05}.
For example, \citet{ume99} calculated evolutions
of 3--9$~M_\sun$ stars for the metallicity of $Z=0.001$--0.03 from
ZAMS until the second dredge-up stage, that is,
until the early AGB phase.  They showed that an upper limit of
C+O core mass is $M_{\rm C+O}= 1.07~M_\sun$ for the initial MS mass
of $M_i = 8.1~M_\sun$ ($Z=0.02$) and for $M_i = 7.1~M_\sun$ ($Z=0.004$).  
If this 1.07$~M_\sun$ upper limit is applied to all the C+O WDs,
both the SD and DD scenarios hardly explain super-luminous SNe~Ia
such as SN~2007if/SN~2009dc.  In this section, we discuss possible
evolutionary paths toward our candidates.

\subsection{Metallicity Effects of C+O Core Masses}
\label{metallicity_effect_co_core}
  In close binaries, the primary star fills its
Roche lobe before it evolves to an AGB star.  In relatively
wide binaries, a star will evolve further beyond the early AGB
phase (the second dredge-up stage).  In the AGB phase,
hydrogen shell-burning produces helium and that
helium piles up on the C+O core.  The helium intermittently ignites
to trigger a shell-flash, making a thermal pulse phase (TP-AGB phase).
The C+O core mass would increase further beyond $1.07~M_\sun$
during the TP-AGB phase until all hydrogen envelope is lost by winds
(or stripped away by the companion star).  
How much mass can be added to the C+O core during the TP-AGB phase ?
We roughly assume an average increasing rate of the core mass from
hydrogen shell-burning rate, $\dot M_{\rm c} \sim 5 \times
10^{-7} M_\sun$~yr$^{-1}$, during the TP-AGB phase
(see Equation [\ref{steady_accretion_wind}]).  On the other
hand, the dust-driven wind mass-loss rates are observationally estimated to
be about $\dot M_{\rm w} \sim 3 \times 10^{-5} M_\sun$~yr$^{-1}$
\citep{deb10}, where the mass-loss rate levels off for stars with
longer pulsation periods of $\gtrsim 850$ days.
Therefore, the ratio of the wind mass-loss rate to the mass-accretion
rate is about $\zeta \equiv \dot M_{\rm w}/\dot M_{\rm c}
\sim 60$.  This simply suggests that the upper limit
for the increase in the core mass is about
$\Delta M_{\rm c} \approx (M_{1,i} - M_{\rm 1,c}) /\zeta 
\sim 6 M_\sun/60 \sim 0.1 ~M_\sun$, where $M_{1,i}$ and $M_{\rm 1,c}$
are the initial mass and the C+O core mass of the primary, respectively.
Furthermore, recent models of massive AGB stars show very efficient third
dredge-up and hot-bottom burning \citep[e.g.,][for a recent
review]{her05}, both of which reduce the core growth.
Therefore the core of massive AGB stars seems to hardly grow
\citep[see, e.g.,][for initial-final mass relation of single stars]{wei00}.


\begin{figure*}
\epsscale{0.6}
\plotone{f7.epsi}
\caption{
(a) Final core masses of AGB stars against metallicity ($Z$).
Here $M_{\rm c,final}$ is the final core mass
of AGB evolutions in solar mass units.
Different symbols denote different authors' results.
Red symbols indicate C+O cores while blue O+Ne+Mg cores.
Circles: C+O/O+Ne+Mg cores taken from \citet{sie10}.
Triangles: C+O cores taken from \citet{wei09}.
Stars: C+O/O+Ne+Mg cores taken from \citet{ven09} and
\citet{ven11}.
Squares: C+O cores taken from \citet{kar10}.
The red dashed line indicates $1.07~M_\sun$ of C+O cores.
The blue dash-dotted line denotes a line below which
C+O cores may exist for lower metallicities of $Z \le 0.002$
estimated from the results of \citet{sie10}.
(b) Core-mass increase, $\Delta M_{\rm c}$, after the first thermal
pulse until the end of AGB evolution against metallicity.
Symbols are the same as those in (a).
\label{co_one_core_growth}}
\end{figure*}

Many authors have calculated the evolutions
of TP-AGB stars, but full calculation through thermal pulses
is extremely time-consuming.  Thus the evolution has been often
followed only through a small number of pulses and then replaced
by synthetic models towards the end of the AGB.  Very recently, a few full
AGB calculations appeared including a number of thermal pulses
until the end of the AGB phase \citep[e.g.,][]{kar10,sie10,ven09,wei09}.
The resultant WD mass, however, depends strongly on the adopted
mass-loss law, extra mixing (overshooting of convection),
hot bottom burning, etc.  Different laws gave different results.

Figure \ref{co_one_core_growth} shows such diverse of resulting 
WD masses through full TP-AGB calculations.
All red filled symbols indicate the C+O core masses after the AGB evolution
for the initial ZAMS mass of $M_i \ge 6~M_\sun$.  
On the other hand, blue open symbols denote the O+Ne+Mg core masses
at the end of the AGB evolution for the initial ZAMS mass of
$M_i \ge 6.5~M_\sun$.  Different symbols correspond to different authors:
circles \citep{sie10}, squares \citep{kar10},  triangles \citep{wei09},
and stars \citep{ven09,ven11}.

Ventura \& D'Antona's (2009, 2011) results
show $\sim 1.07~M_\sun$ as an upper limit of C+O WDs for a wide 
range of metallicities, i.e., for $Z=0.0001$--0.004.  On the other hand,
Siess's (2010) results show a considerable increase in the C+O core
mass for very low metallicities such as $Z \lesssim 0.002$.
The main differences probably come from the different treatments of
wind mass-loss during the TP-AGB and possibly extra-mixing,
because \citet{sie10} assumed Schwarzschild criterion of
convection, which prevents the third dredge-up, and also assumed
Vassiliadis \& Wood's (1993) mass-loss rate, which makes
the duration of the TP-AGB phase much longer than that of
Bl\"oker's (1995) formalism adopted by \citet{ven09}. 
\citet{ven11} showed that Vassiliadis \& Wood's formalism gives
about 9 times more thermal pulses than Bl\"oker's formulation
for the same initial mass of $7~M_\sun$ ($Z=0.001$), i.e., 348 vs. 38,
because of weaker wind mass-loss rates and thus a longer duration of
the TP-AGB phase.  If Vassiliadis \& Wood's one is the case,
we have C+O WDs more massive than $1.1~M_\sun$ for lower metallicities,
which can open an evolutionary route to super-Chandrasekhar mass
SNe~Ia as massive as $2.3$--$2.7~M_\sun$.

These two formalisms by \citet{vas93} and \citet{blo95}
do not explicitly include metallicity effects.  Observationally
there are some suggestions on metallicity effects of AGB winds.
Local galaxies with low $Z$ show some evidence for a deficiency 
in the number of planetary nebulae (PNe) at [Fe/H]$\lesssim -1$
\citep{mag03,zij04}.
The deficiency of PNe suggests that the mass-loss rates are not
as high as those observed in metallicities 
of [Fe/H]$\gtrsim -1$. In other words,  a superwind does not blow
in lower metallicity environment \citep{zij04}.  These changes of
behavior occur at [Fe/H]$\approx -1$
\citep[see also][for a theoretical discussion on this metallicity
effect]{bow91}, suggesting that,
in lower $Z$~ ([Fe/H]$\lesssim -1$) environments, C+O WDs
possibly become more massive than those in higher $Z$~ 
([Fe/H]$\gtrsim -1$) environments.

From these circumstances, we expect that AGB stars produce C+O WDs
with mass higher than 1.1$~M_\sun$ in lower metallicities of
$Z \lesssim 0.002$. Since the mass loss is relatively weak in lower
metallicity AGB stars, the C+O core is likely to grow as massive as 
$1.2~M_\sun$ or more before the binary undergoes a common envelope
evolution.  In such a case, the maximum C+O WD mass can reach
2.3--$2.7~M_\sun$ by accretion (see Figure \ref{maximum_wd_mass_c10}a).
Observational suggestions on massive C+O WDs and low metallicity
environments of super-$M_{\rm Ch}$ SNe~Ia will be discussed in Section
\ref{metallicity_effect_sn1a}.

Finally, we comment on Nomoto \& Kondo's (1991) 
upper limit for the initial C+O WD mass:
$M_{\rm WD,0} \lesssim 1.2~M_\sun$ for SN~Ia explosions
\citep[see Figure 4 of][]{nom91}.
This upper limit, however, does not apply to our super-Chandrasekhar mass
WDs.  This upper limit is valid only when
the central part of the WD stays cold so that central carbon burning
leads to a collapse instead of explosion.
In the normal case of Chandrasekhar mass SNe~Ia, compressional heat due
to mass-accretion does not reach the central part in a short
timescale, e.g., that a WD grows from 1.3 to 1.38$~M_\sun$.
On the other hand, for super-Chandrasekhar mass WDs,
the timescale of mass-accretion becomes much 
(a few to several times) longer because it takes more than a few
to several times for a WD to grow, e.g., from 1.3 to 1.8$~M_\sun$.
At the time of central carbon ignition, the central part has been
heated up, thus central carbon burning leads to an SN~Ia explosion.


\begin{figure}
\epsscale{1.15}
\plotone{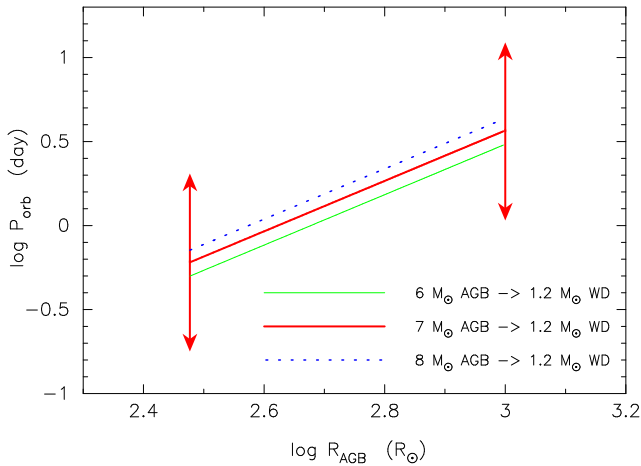}
\caption{
Final orbital periods against the radii of AGB stars just
before the common envelope evolution.
We assume that the AGB star fills its Roche lobe when the C+O
core grows to $1.2~M_\sun$ and a common envelope evolution
starts.  We plot three cases of mass of the AGB star, i.e.,
6 ({\it green thin}), 7 ({\it red thick}), and $8~M_\sun$
({\it blue dotted}).  Vertical arrows indicate upper/lower
dispersions of the factor $\alpha \lambda$ in Equation
(\ref{ce_shrink}). (See Equations [\ref{ce_lambda}]
and [\ref{ce_alpha}]).
\label{xm7m4_shrink}}
\end{figure}

\subsection{First Common Envelope Evolution and Binary Orbital Period}
\label{first_common_envelope}
Now we examine the condition for the initial orbital period
($P_{\rm orb,0}=0.5$--1.5 days) to be realized when an AGB star
with a core mass of $M_{\rm C+O}=1.1$--1.2$~M_\sun$
undergoes a common envelope evolution.
\citet{dem11} analyzed the post common envelope evolution systems
with the relation of
\begin{equation}
-G{{M_c (M_e/2+M_c)} \over {\lambda R}} = -\alpha G \left[
{{M_c M_2} \over {2 a_f}} - {{(M_c+M_e)M_2} \over {2 a_i}} \right],
\label{ce_shrink}
\end{equation}
where $M_c$ and $M_e$ are the giant primary's core mass
and the primary's envelope mass, $a_f$ and $a_i$ the
final and initial separation of the binary, $M_2$ the secondary's
mass, which is not changed during the common envelope evolution,
$\lambda$ is the number of order unity which depends on the mass
distribution of the primary's envelope ($\lambda R$ is 
effectively the mass-weighted mean radius of the envelope),
and $\alpha$ the efficiency parameter of order unity.
They obtained $\lambda$ and $\alpha$ for AGB stars as
\begin{equation}
\lambda_{\rm AGB}= (0.237 \pm 0.021) + (0.032 \pm 0.006)
\times {{M_{\rm MS}} \over {M_\sun}},
\label{ce_lambda}
\end{equation}
\begin{equation}
\log \alpha = (-1.4 \pm 0.2)  - (1.2 \pm 0.2) \log {M_2 
\over M_1},
\label{ce_alpha}
\end{equation} 
where $M_{\rm MS}$ is the main-sequence mass of the AGB star, and
$M_1$ and $M_2$ are the primary and secondary masses
just before the common envelope evolution, respectively.
We do not know the exact radius of TP-AGB stars just before
the common envelope evolution, so we roughly assume 
the radius of $R \sim 300$--$1000~R_\sun$
\citep[see, e.g.,Figures 1 and 2 of][]{web07}, and the C+O core
mass of $1.2~M_\sun$, and $M_1=6$, 7, and $8~M_\sun$, and 
$M_2=4~M_\sun$.  Then we have $a_i \sim 700$--2200$~R_\sun$.
Substituting these values into Equations
(\ref{ce_shrink}), (\ref{ce_lambda}), and (\ref{ce_alpha}),
we estimate the final separation of the binary.
Figure \ref{xm7m4_shrink} shows the orbital period of
the binary against the radius of the AGB star just before
the common envelope evolution.  We obtain rather short separations
and orbital periods of $a_f \sim 5$--$17~R_\sun$ and
$P_{\rm orb,0}\sim 0.6$--3.6 days for $M_{1,i}=7~M_\sun$, 
$M_{\rm 1,C+O}= M_{\rm WD,0}= 1.2~M_\sun$, and $M_{2,i}=4~M_\sun$
Incorporating ambiguity of Equations
(\ref{ce_shrink})--(\ref{ce_alpha}) in common envelope evolution,
the resultant binary periods
lie possibly in the region of 0.5--1.5 days, which is consistent
with our requirement for super-Chandrasekhar mass SNe~Ia of
$M_{\rm WD} \sim 2.3$--$2.7~M_\sun$.

\section{Discussion}

\subsection{Comparison with Other SD Models}
\citet{che09} calculated binary evolutions to SNe~Ia without including
mass-stripping effect.  Our results can be compared with their cases
of very short WD wind phases because of small mass-stripping effects.
\citet{che09} obtained the maximum WD mass of $1.75~M_\sun$
at $t\sim1.5$~Myr (after the companion fills its Roche lobe) for
$M_{\rm WD,0}=1.2~M_\sun$, $M_{2,0}=2.5~M_\sun$,
and $P_{\rm orb,0}=1$~day (with $c_1=0$ in our definition).
We obtained $M_{\rm WD}=1.87$, 1.88, $1.89~M_\sun$ at $t\sim1.5$~Myr
for the same initial model with $c_1=1$, 3, and 10, respectively.
Our results are consistent with theirs.  The small difference
of $\sim 0.1~M_\sun$ might stem from the difference in the 
mass-accretion rates onto the WD;  Chen \& Li assumed smaller
accretion rates than ours by taking into account centrifugal force,
thus they get a smaller WD mass.

\subsection{Detectability of SD signatures}
In a recent review, \citet{how11} listed four major
arguments strongly against the existing SD scenarios:
(1) very few supersoft X-ray sources have been found in galaxies;
(2) no shock interaction between ejecta and a companion star has been seen;
(3) no hydrogen lines have been detected;
(4) no explanation of super-Chandrasekhar mass type Ia supernovae has been
given.

As for the first argument, although being not directly related to
the super-Chandrasekhar mass models, \citet{gil10} claimed
that observed soft X-ray fluxes of early type galaxies
contradict the SD scenario based on simple assumptions.
However, \citet{hkn10a} showed that realistic evolutionary models of
the SD systems spend a large fraction of the life time in the optically
thick wind phase and the recurrent nova phase rather than the supersoft
X-ray source (SSS) phase.
Thus the SSS phase lasts only for a few hundred thousand years.  This
is by a factor of $\sim 10$ shorter than those adopted by \citet{gil10}
where the SN~Ia progenitor WD was assumed to spend
most of its life as a SSS.
The X-ray luminosity of the SSS has a large uncertainty because of the
uncertain atmospheric model of mass-accreting WDs and absorption of
soft X-rays by the companion star's cool wind material.  
Thus adopting an average of the observed fluxes of existing symbiotic
SSSs, \citet{hkn10a} showed that the observed X-ray fluxes in early
type galaxies obtained by \citet{gil10} are rather consistent
with the SD scenario.

As for the second (and third) points, \citet{how11} claimed that
there should be some observational evidence of shock interaction
between ejecta and the companion star \citep{kas10} such as distortions
in early-time light curves of SNe~Ia, especially for a RG companion,
but no such signatures have been observed.  This is problematic for
the SD scenario of WD+RG systems \citep[][see, however,
a different (rather contradict) result by \citet{gan11}]{hay10, bia11}.
These shocking signatures predicted in \citet{kas10} are
based on the assumption that the companion fills its Roche lobe.
However, if the binary separation, $a$, is much larger than 
the companion radius, $R$, i.e., $a \gg R$, the solid angle
subtended by the companion would be much smaller, and so would
be the effect of shocking. \citet{jus11} and \citet{dis11}
argued that the donor star in the SD scenario
might shrink rapidly before the WD explosion, because it would exhaust
its hydrogen-rich envelope during a long spinning down time of
the rapidly rotating WD until the SN~Ia explosion.
In such a case, the companion star
would be much smaller than its Roche lobe, reducing the shocking
signature, which also explains the lack of hydrogen 
in the spectra of SNe~Ia.

In our super-Chandrasekhar mass SN~Ia models, the SN~Ia explosion occurs
after the mass of the main-sequence (MS) secondary decreases to
as small as 1.5--1.8 $M_\sun$ (see, e.g., two evolution models in 
Figure \ref{zregevl10_strip_ms_rg_wd11}).
Impact with such a low mass MS companion would not produce
a large distortion in early-time SN~Ia light curves
\citep[see, e.g., Figure 3 of][]{kas10},
so no detection of distortion is not a problem at least
for our super-Chandrasekhar mass models.

As for the third disfavor point, \citet{how11} claimed that
if WDs grow to become SNe~Ia via hydrogen accretion,
there should be some indication of hydrogen in some form,
but no such indication has been found, e.g.,
in radio emission \citep{pan06} nor in stripped ejecta from
the companion \citep{mar00}.
\citet{jus11} and \citet{dis11} argued that the
donor star in the SD scenario would shrink rapidly to become a helium
white dwarf before the WD explosion, having exhausted its hydrogen-rich
envelope.  Therefore, no detection of hydrogen would naturally be expected.

In our super-Chandrasekhar mass SN~Ia models, the SN~Ia explosion occurs
after the mass of the main-sequence (MS) secondary decreases
to as small as $\sim 2.0~M_\sun$ (see, e.g., model a in 
Figure \ref{spindown_evol_ms_wd_c1_10}).
Impact with such a low mass MS companion would lead to a strip of
hydrogen-rich matter as massive as $\sim 0.1~M_\sun$ (luminous
super-Chandrasekhar mass SNe~Ia), producing hydrogen lines
in a later phase as shown in \citet{leo07}.

As for the fourth (last) point, which is the theme of this paper,
we have already shown the possible cases where the WD mass increases
up to $\sim 2.4~M_\sun$ based on our SD scenario.
Thus, we are able to resolve the four major arguments raised
by \citet{how11} against the SD scenario.

\subsection{Young Progenitors in Low Metallicity Environments} 
\label{metallicity_effect_sn1a}
Our calculations show that super-Chandrasekhar mass SNe~Ia come
preferentially from a pair of a massive C+O WD ($\gtrsim 1~M_\sun$)
and a $4-5~M_\sun$ main-sequence star, thus indicating a rather
younger population than several hundred Myr or so.
In Section \ref{metallicity_effect_co_core}, we further show that
very luminous super-Chandrasekhar mass SNe~Ia such as
SN~2007if/SN~2009dc
are preferentially born in low metallicity environment because
more massive initial C+O WDs  ($\sim 1.1$--$1.2~M_\sun$) are
required.  Here we discuss these conditions in view of observational
aspects.

Recent observation of the SN~2007if host galaxy suggests a low
metallicity environment of $Z\sim Z_\sun/9$--$Z_\sun/5$ as well as
a young population of the progenitor from the host galaxy turn-off
mass of $M \sim 4.6^{+2.6}_{-1.4} ~M_\sun$ \citep{chi11}, which
is consistent with our luminous super-Chandrasekhar mass
model that can attain the maximum WD mass at 4--5~$M_\sun$ 
in our binary evolutions (Figure \ref{maximum_wd_mass_c10}). 
We could have super-Chandrasekhar mass models even in such a
low metallicity environment.
Although \citet{kob98} showed that SNe~Ia may not frequently occur
in low metallicity environments (say, $Z \lesssim Z_\sun/10$)
because optically thick winds are weak, we here point out that
the metallicity of the SN~2007if host galaxy is still large enough
for massive WDs (with the initial mass of $M_{\rm WD,0}=1.1$--$1.2~M_\sun$)
to blow strong winds (see Figure \ref{metal_new}).
Even for the metallicity of Population II
($Z=0.001$), massive WDs ($M_{\rm WD,0} \ge 1.1~M_\sun$)
can blow strong winds and, as a result, the stripping effect works.  
Thus, in such a low metallicity environment, 
we expect (super-Chandrasekhar mass) SN~Ia explosions in our SD scenario.

Young population of the progenitor was also suggested for SN~2009dc.
\citet{tau11} suggested that UGC~10063 (suspected host galaxy
of SN~2009dc if the supernova is in the edge of the tidal tail
toward the galaxy UGC~10064) underwent a strong burst of star formation
a few hundred Myr before the explosion of SN~2009dc
\citep[see also Figure 14 of][for images and position of UGC~10063,
UGC~10064, and SN~2009dc]{sil11}.
Our favorite model has an MS companion of
mass $\sim4$--5~$M_\sun$ .  A 4--5$~M_\sun$ MS evolves to 
the end of central hydrogen burning a few hundred Myr
after its born.  Thus our model is consistent with these circumstances.

From these observational aspects, we may propose that luminous
super-Chandrasekhar mass SNe~Ia come preferentially from a low
metallicity environment.  This is
because the AGB wind (or superwind) is relatively weak 
for $Z \lesssim 0.002 \sim Z_\sun/10$ so that the C+O WD can
grow up to $\sim 1.2~M_\sun$ before the first common envelope evolution.
On the other hand, the WD accretion (optically thick) wind
is still strong enough for a $\sim 1.2~M_\sun$ WD,
if the metallicity is larger than $Z \gtrsim 0.0002 \sim Z_\sun/100$.
Considering these rare circumstances
(a narrow range of WD masses and a low metallicity environment), 
we expect rather rare events of super-Chandrasekhar mass SNe~Ia.

\section{Conclusions}
Recent observations of Type~Ia supernovae (SNe~Ia) suggest 
that some of the progenitor white dwarfs (WDs) had masses
up to 2.4--$2.8~M_\sun$, which are highly exceeding the Chandrasekhar
mass limit.  Introducing three binary evolution processes into our
binary evolution code, i.e.,
optically thick winds from mass-accreting WDs, mass-stripping
from the binary companion star by the WD winds,
and WDs being supported by differential rotation,
we have calculated new SD evolutionary models of SN~Ia
progenitors and found following main results:

{\bf 1.}~Since the mass-stripping attenuates the mass transfer from
the companion to the WD, which prevents a binary even with a massive
main-sequence companion of $\sim 4$--$5~M_\sun$ from being merged.
As a result, the WD mass can increase
by accretion up to 2.3~(2.7)$~M_\sun$
from the initial value of 1.1~(1.2)$~M_\sun$.
The newly obtained results satisfy the requirements of
recent high luminosity SNe~Ia such as SN~2007if and SN~2009dc.

{\bf 2.}~There are three characteristic mass ranges of exploding WDs
depending on the secular instability and rotation law:
(i) $M_{\rm WD}>2.4~M_\sun$ triggered by a secular instability 
at $T/|W|=0.14$, (ii) $M_{\rm WD}=1.5$--2.4$~M_\sun$ supported by
differential rotation, and (iii) $M_{\rm WD}=1.38$--1.5$~M_\sun$
supported by rigid rotation.  These three cases are explained 
as follows:

{\bf 3.}~If the initial C+O WD is as massive as
$\sim 1.1$--$1.2~M_\sun$, $T/|W|$ exceeds 0.14 when the WD mass
reaches $M_{\rm WD} \approx 2.4~M_\sun$.  We expect an SN~Ia explosion
soon after a secular instability sets in at $M_{\rm WD} \approx 2.4~M_\sun$.
We regard that this kind of binary evolution corresponds to 
very luminous super-Chandrasekhar mass SNe~Ia such as
SN~2007if/SN~2009dc.  These SNe~Ia must be young
(younger than a few hundred Myr).

{\bf 4.}~If the initial C+O WD mass is $\sim 0.9$--$1.1~M_\sun$,
$T/|W|$ does not exceed 0.14 before $\dot M_1$ becomes smaller than
$\dot M_{\rm b} = 3 \times 10^{-7}~M_\sun$~yr$^{-1}$,
where we stop our binary evolution.  Loss or redistribution of
angular momentum leads to an increase in the central density
until carbon is ignited to induce the WD explosion
as an SN~Ia.  This waiting time for an SN~Ia depends entirely
on the timescale of angular momentum loss/redistribution.
It is shorter than $10^8$--$10^9$~yr (angular momentum
redistribution by the Eddington-Sweet circulation in WDs).
Therefore we suggest that these SNe~Ia
are relatively young, and correspond to the ``prompt'' component.  

{\bf 5.}~If the initial C+O WD mass is $\sim 0.7$--$0.8~M_\sun$,
it reaches $M_{\rm WD}\approx 1.38$--$1.5~M_\sun$, which can be
supported by rigid rotation.

\acknowledgments
This research has been supported in part by the Grant-in-Aid for
Scientific Research of the Japan Society for the Promotion of
Science (20540227, 22012003, 22540254, 23105705, 23540262)
and by World Premier International
Research Center Initiative, MEXT, Japan.

\end{document}